\documentclass{edp-jp4}
\usepackage{graphicx}
\begin{document}
\title{Complete synchronization of convective patterns between Gray--Scott systems}
\author{Gonzalo Iz\'us$^*$}\address{Departamento de F\'{\i}sica, Facultad de Ciencias Exactas y Naturales, Universidad Nacional de Mar del Plata,\\De\'an Funes 3350, 7600 Mar del Plata, Argentina.}
\author{Roberto Deza$^+$}\sameaddress{1}
\author{Luis Bernal$^*$}\sameaddress{1}
\author{Vicente P\'erez-Villar}\address{Grupo de F\'{\i}sica Non Lineal, Facultade de F\'{\i}sica, Universidade de Santiago de Compostela,\\E-15782 Santiago de Compostela, Spain.}
\maketitle
\noindent$^*$ Member of CONICET, Argentina.\\$^+$ Invited talk at the Workshop on ``\emph{Complex Systems: New Trends and Expectations}'', Santander (Spain) 5--9 June 2006.
\begin{abstract}
Two identical 1D autocatalytic systems with Gray--Scott kinetics---driven towards convectively unstable regimes and submitted to independent spatiotemporal Gaussian white noises---are coupled unidirectionally, but otherwise linearly. Numerical simulation then reveals that (even when perturbed by noise) the slave system replicates the convective patterns arising in the master one to a very high degree of precision, as indicated by several measures of synchronization.
\end{abstract}
\section{Introduction}\label{intro}
The beautiful talk in this Workshop by J\"urgen Kurths \cite{Kurths} (as well as others dealing to some extent with the subject) exempts us to introduce the field of synchronization at large. Thus, hereafter we shall restrict our scope to the less explored subfield of the synchronization between continuous systems \cite{Parmananda,Kocarev,Junge,Boccaletti}---concretely, to non-delayed synchronization between systems of stochastic partial differential equations. In particular, a topic that has been hardly addressed is the synchronization between noise-sustained structures (NSS) in systems undergoing a convective instability \cite{opo2}.

A \emph{convectively} unstable regime is characterized by the fact that local perturbations to the steady state are advected more rapidly than their spreading rate \cite{Deissler}. When seen in a Lagrangian framework, the system is unstable; from an Eulerian description, however, perturbations are ``washed out by the flow''. Macroscopic patterns named \emph{noise sustained structures} (NSS) emerge in this regime if noise is present at all times. It is through dynamical amplification of random fluctuations that the system is driven out of its linearly unstable steady state towards the state sustaining NSS. Thus, if noise (or any external deterministic forcing) were not present, nonequilibrium structures could not arise. In fluid dynamics the NSS are a spatial macroscopic manifestation of amplified thermal fluctuations.

NSS have been observed in fluid convection experiments (both in open flow configuration \cite{ahlers} and Taylor--Couette flows \cite{Bab,Tsam}), and their precursors have been also observed in nematic liquid crystals \cite{Rehberg}. They have also been numerically shown to exist in optical systems \cite{opo2,Marco} (driven in this case by quantum noise) and recently, in a model autocatalytic chemical reaction---the Gray--Scott (GS) model---taking place in a differential-flow reactor \cite{GS}.

The paper is organized as follows: In Sec.\ \ref{sec:model} a brief sketch is made of the GS model, and the rationale and features of the chosen master--slave coupling are pointed out. Section\ \ref{sec:numsim} introduces the details of the numerical integration scheme and discusses the features of the NSS arising in the uncoupled systems. Section \ref{sec:sync1} is devoted to a fairly thorough numerical characterization of the replication of NSS through complete synchronization. In particular, the behavior of the synchronization measures as functions of the parameters in the model is studied, and a numerical estimation is made of the robustness of the phenomenon. Finally, the main conclusions are summarized in Sec.\ \ref{sec:concl}.
\section{The model}\label{sec:model}
The GS model proposes three steps for the conversion of the precursor species $P$ into the inert product $C$ \begin{eqnarray*}
P&\stackrel{k_0}{\rightarrow}&A\\
A+2B&\stackrel{k_1}{\rightarrow}&3B\\
B&\stackrel{k_2}{\rightarrow}&C
\end{eqnarray*}
The intermediate step has cubic autocatalytic kinetics.

In the case we consider, the reaction takes place in a differential-flow reactor where $A$ is immobilized, whereas $B$ is free to diffuse and is also advected by the flow. Moreover, the reaction is maintained out of equilibrium by keeping the concentration of $P$ constant ($p=p_0$) and that of $C$ zero ($c=0$). Hence, the present model describes the irreversible decay of $P$ towards a product $C$ that is immediately removed from the reactor.

After scaling concentrations by $(k_2/k_1)^{1/2}$, time by $k_2^{-1}$ and length by $(D_B/k_2)^{1/2}$, the rate equations for system 1 read
\begin{eqnarray}\label{eq:master1}
\frac{\partial a_1}{\partial t}&=&\mu-a_1\,b_1^2+\sqrt{\sigma_1}\,
\xi_1(\mathbf{r},t),\nonumber\\
\frac{\partial b_1}{\partial t}&=&\nabla^2b_1-\phi\,\nabla b_1-b_1+a_1\,b_1^2,
\end{eqnarray}
where $\mu$ stands for the scaled version of $k_0p_0$ and $\phi$ for that of the fluid velocity $v$. The real Gaussian noise $\xi_1$ in the rate equation for $a_1$---with zero mean, variance $\sigma_1$, and delta-correlated in space and time---accounts for fluctuations (either thermal or in $p_0$).

For $\mu>1$, the uniform steady state $(a_1=\mu^{-1},b_1=\mu)$ becomes convectively unstable at some $\phi_c(\mu)$, yielding to traveling waves with $\pm q_c$. Further details are found in \cite{GS} and references therein.

Now we assume that system 1 drives another system (called thereafter system 2 and lying in a second differential-flow reactor) in a master--slave configuration. System 2 has the same values of $\mu$ and $\phi$ but its $A$--component is submitted to a spatiotemporal Gaussian white noise $\xi_2(\mathbf{r},t)$ with a possibly different noise variance $\sigma_2$:
\begin{eqnarray}\label{eq:master2}
\frac{\partial a_2}{\partial t}&=&\mu-a_2\,b_2^2+\sqrt{\sigma_2}\,\xi_2(\mathbf{r},t),\nonumber \\
\frac{\partial b_2}{\partial t}&=&\nabla^2b_2-\phi\,\nabla b_2-b_2+a_2\,b_2^2+\epsilon\,(b_1-b_2).
\end{eqnarray}
$\epsilon$ denotes the strength of the unidirectional \emph{linear} coupling between both reactions. Besides being the simplest coupling that enables synchronization, it facilitates an approach to the stability analysis of the synchronization manifold \cite{IDB}.
\section{Numerical simulation and unsynchronized NSS}\label{sec:numsim}
We shall restrict hereafter to the 1D case (the specificities found in higher spatial dimensions will be published elsewhere \cite{IDB}). Equations (\ref{eq:master1}) and (\ref{eq:master2}) have been integrated using an Euler stochastic scheme in a grid of $16384$ sample points with a grid space $\Delta x=0.1$ and time step $\Delta t=0.0001$. The parameters have been chosen as $\mu=2.0$, $\phi=9.5$, $\sigma_1=10^{-7}$. For each system, Dirichlet BC is assumed at the inlet of the reaction domain [$a_i(0,t)=\mu^{-1},\,b_i(0,t)=\mu,\,(i=1,2)$] and Neumann BC at the outlet ($x=L$). The length $L$ is chosen in such a way that spatiotemporal patterns develop well before they reach the outlet.

For $\epsilon=0$, Eqs.\ (\ref{eq:master1})--(\ref{eq:master2}) describe two uncoupled reactions, identical with regard to the deterministic parameters but submitted to independent spatiotemporal noises which produce non-correlated NSS in both systems. These patterns have been characterized in Ref.\ \cite{GS}: they are dynamical structures that drift with the flow, disappearing on the right, whereas new wave excitations are continuously regenerated by dynamical amplification of noise.
\begin{figure}
\begin{center}
\includegraphics[width=3.2in,bb= 54pt 360pt 558pt 720pt]{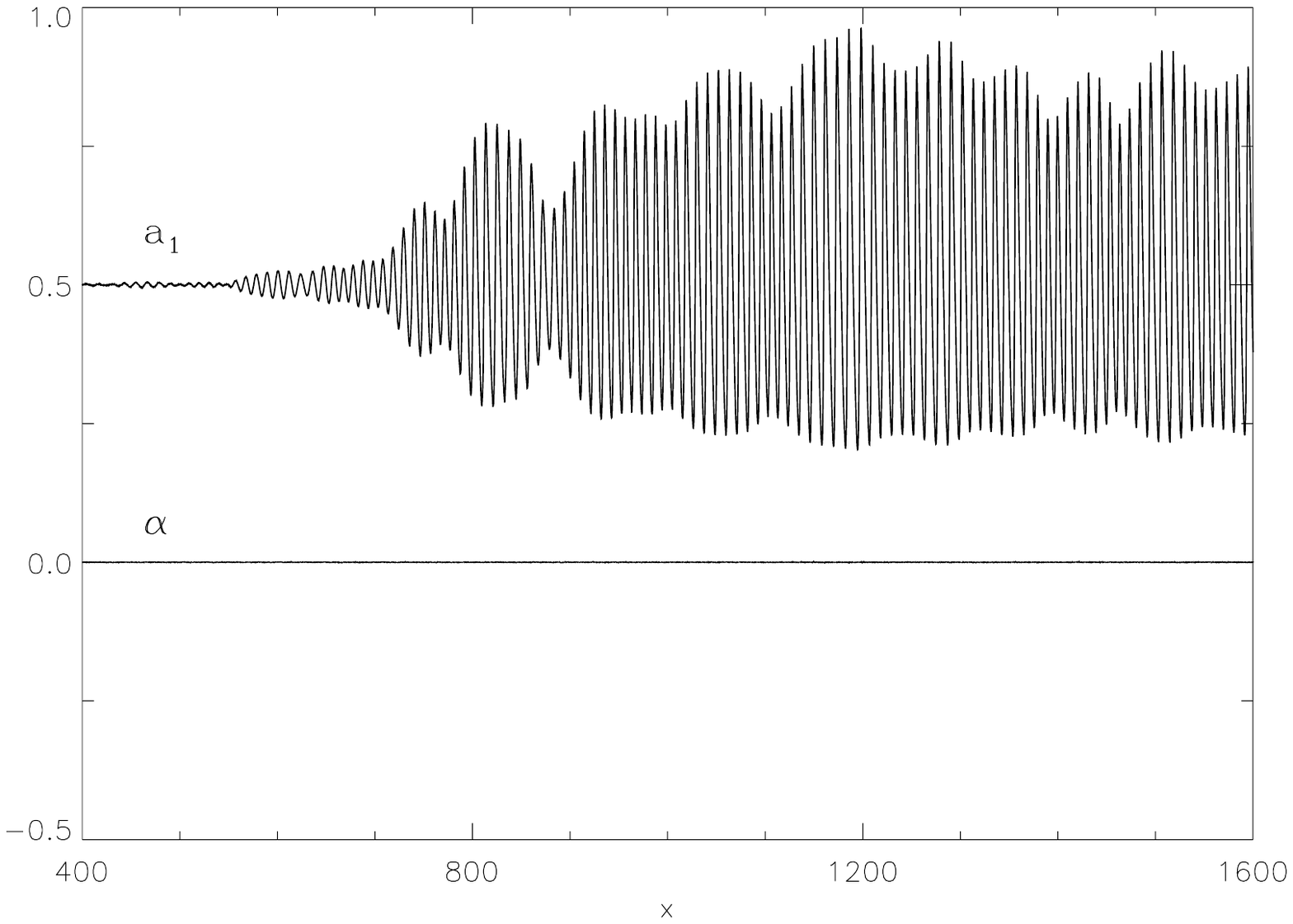}\\
\includegraphics[width=3.2in,bb= 54pt 360pt 558pt 720pt]{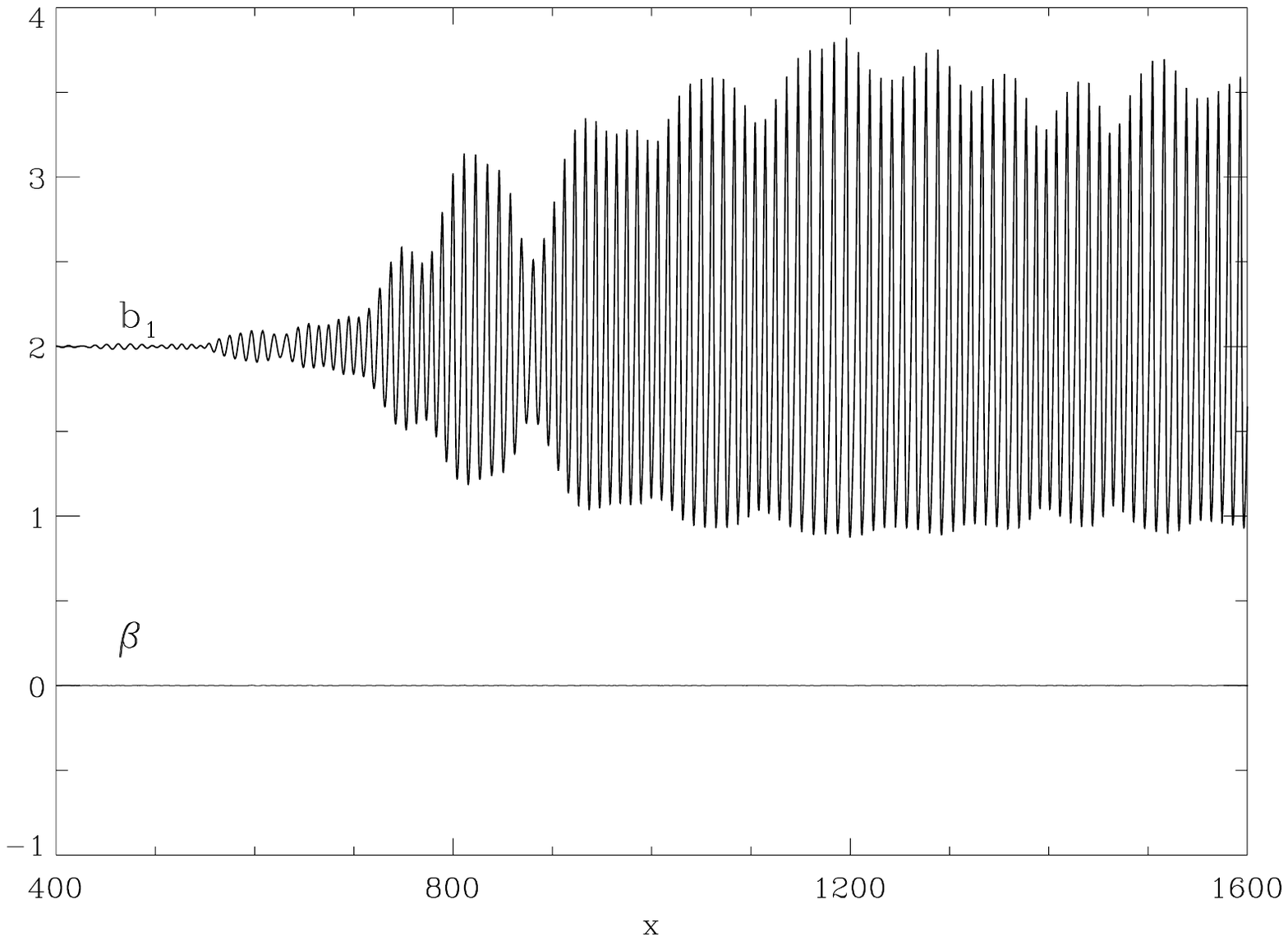}\\
\end{center}
\caption{A typical snapshot in the simulation of Eqs.\ (\ref{eq:master1}) for
$\epsilon=0.1$.
(a) $a_1$ vs $x$; (b) $b_1$ vs $x$. Also shown---and in the same corresponding scales---are the deviation fields  $\alpha$ and $\beta$. The remaining parameters are $\mu=2.0$, $\phi=9.5$ and $\sigma_{1,2}=10^{-7}$.}\label{fig:1}
\end{figure}
\section{Synchronization of noise-sustained structures}\label{sec:sync1}
When $\epsilon\neq0$, some correlation is expected between the NSS in both systems. A handy measure of correlation for these snapshots are the deviation fields
\begin{eqnarray}\label{eq:dif}
\alpha(x,t)&=&a_1(x,t)-a_2(x,t),\nonumber\\
\beta(x,t)&=&b_1(x,t)-b_2(x,t).
\end{eqnarray}
Figures \ref{fig:1}a and \ref{fig:1}b are respectively snapshots of typical $a_1$ and $b_1$ profiles. The deviation fields are also depicted (in solid lines and in the same scales as $a_1$, $b_1$ respectively) in Figs.\ \ref{fig:1}a and \ref{fig:1}b. The result is surprising, given that system 2 is also submitted to an independent spatiotemporal noise source: In the scales of Figs.\ \ref{fig:1}, \emph{system 2 synchronizes completely to system 1}. In other words, an effective replication of the NSS arising in the first reactor takes place at the second one, and time evolution---even under the influence of noise---does not spoil the high degree of synchronization.

If we regard the system's evolution as a succession of snapshots like those of Figs.\ \ref{fig:1}, natural quantifiers for this phenomenon (as functions of $t$) are the variances of $\alpha$ and $\beta$:
\begin{eqnarray}\label{eq:var}
\sigma_\alpha(t)&\equiv&
\sqrt{\frac{1}{L}\int_0^L[\alpha^2(x,t)-\langle\alpha\rangle^2]\,dx},\nonumber\\
\sigma_\beta(t)&\equiv&\sqrt{\frac{1}{L}\int_0^L [\beta^2(x,t)-\langle\beta\rangle^2]\,dx},
\end{eqnarray}
with $\langle\varphi\rangle_{(t)}\equiv\frac{1}{L}\int_0^L\varphi(x,t)\,dx$ ($\varphi$ stands for $\alpha$ and $\beta$ respectively), and the global synchronization error
\begin{equation}\label{eq:global}
E(t)=\sqrt{\frac{1}{L}\int_0^L[\alpha^2(x,t)+\beta^2(x,t)]\,dx}.
\end{equation}

\begin{figure}
\begin{center}
\includegraphics[width=3.2in,bb= 54pt 360pt 558pt 720pt]{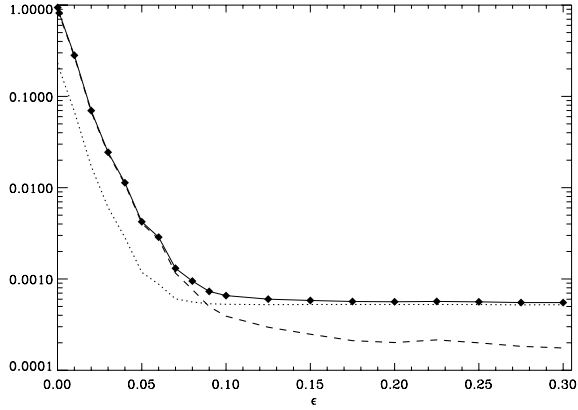}\\
\end{center}
\caption{Synchronization measures as functions of the coupling $\epsilon$: $E$ (solid line), $\sigma_{\alpha}$ (dotted line) and $\sigma_{\beta}$ (dashed line). The remaining parameters are $\mu=2.0$, $\phi=9.5$ and $\sigma_{1,2}=10^{-7}$.}\label{fig:2}
\end{figure}

Figure \ref{fig:2} is a plot of $E$, $\sigma_\alpha$ and $\sigma_\beta$ vs $\epsilon$ for typical realizations (as stated before, the time evolution preserves the degree of synchrony). In the numerical simulation, $\langle\alpha\rangle$ and $\langle\beta\rangle$ remain below $\sim 5 \times 10^{-5}$, i.e., $E^2\sim\sigma_\alpha^2+\sigma_\beta^2$ and basically $E$ accumulates the information of both variances. On the other hand, correlations between the NSS remain during time evolution above $.9999$, indicating a very high degree of structure replication. In other words, the coupling in Eqs.\ (\ref{eq:master2}) synchronizes the whole stochastic processes.

\begin{figure}
\begin{center}
\includegraphics[width=3.2in,bb= 54pt 360pt 558pt 720pt]{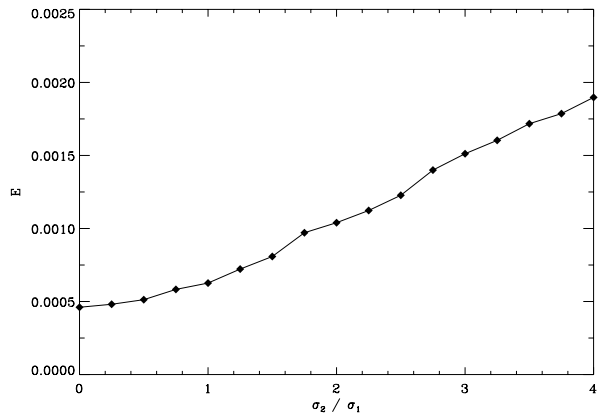}\\
\end{center}
\caption{$E$ vs $\sigma_2/\sigma_1$ for $\sigma_1=10^{-7}$. The $\sigma_{\alpha,\beta}$ variances follow a similar behavior. The remaining parameters are $\mu=2.0$, $\phi=9.5$ and $\epsilon=0.1$.}\label{fig:3}
\end{figure}

A dependence of $\sigma_\alpha$, $\sigma_\beta$ (and thus of $E$) on the noise intensity $\sigma_2$ is to be expected. Figure \ref{fig:3} shows (here $\sigma_1= 10^{-7}$) that this is indeed the case, and maximum synchronization corresponds to $\sigma_2=0$. In other words, only in system 1 does the noise play a constructive role (by pushing the system out of its unstable steady state).

\begin{figure}
\begin{center}
\includegraphics[width=3.2in,bb= 54pt 360pt 558pt 720pt]{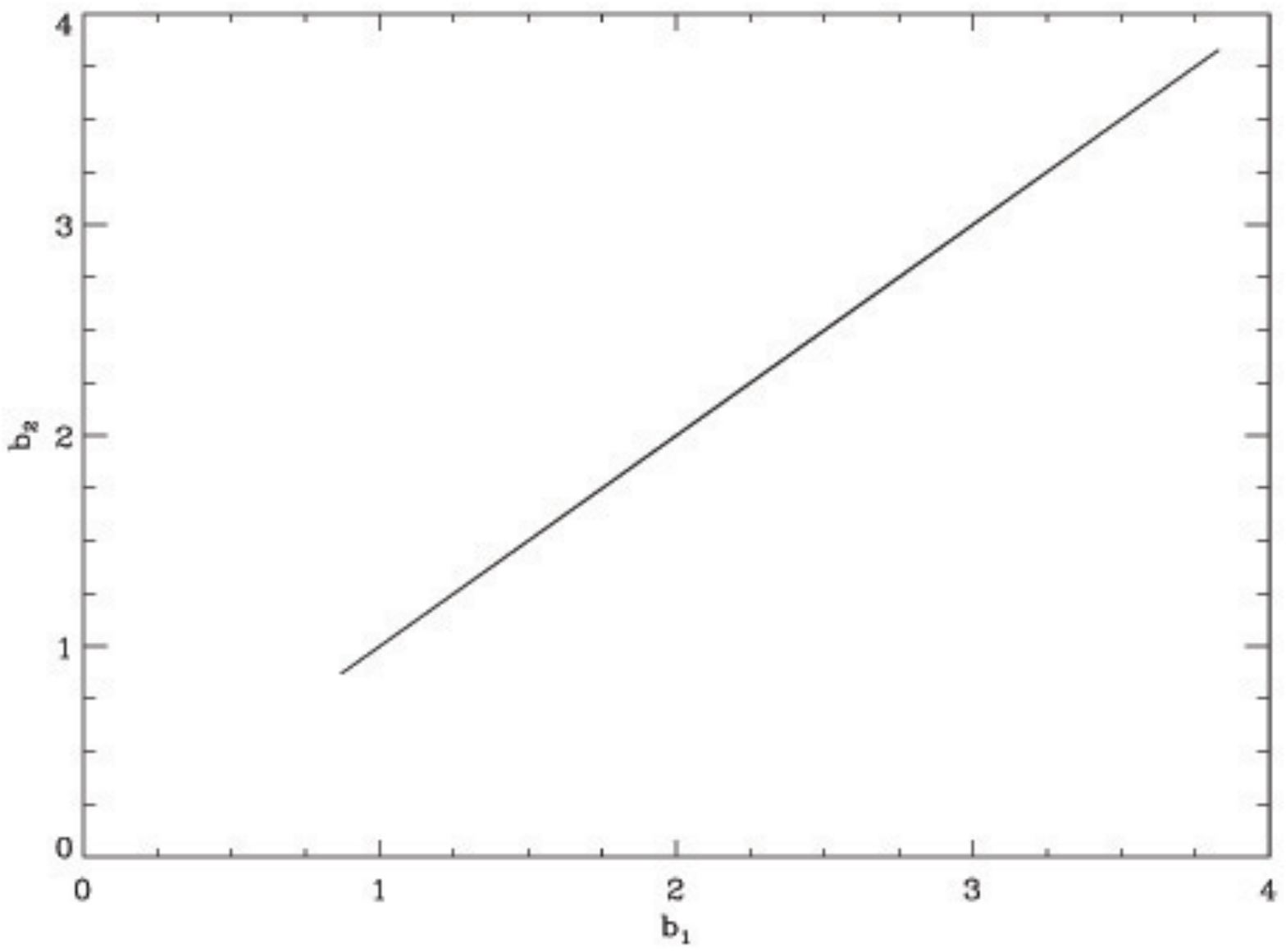}\\
\end{center}
\caption{(a): The $(b_1,b_2)$ phase plane for the complete time history of a numerical realization of Eqs.\ (\ref{eq:master1}) and (\ref{eq:master2}). A similar picture is obtained in the $(a_1,a_2)$ plane. The parameters are $\mu=2.0$, $\phi=9.5$, $\sigma_{1,2}=10^{-7}$ and $\epsilon=0.1$.}\label{fig:4}
\end{figure}

As usual, the synchronization between the stochastic fields $a_1$ and $a_2$ (resp.\ $b_1$ and $b_2$) can also be viewed in the corresponding phase planes. As an illustration, Fig.\ \ref{fig:4} shows the dynamical correlation between $b_1$ and $b_2$ during the complete time history of a numerical simulation.

\begin{figure}
\begin{center}
\includegraphics[width=3.2in,bb= 54pt 360pt 558pt 720pt]{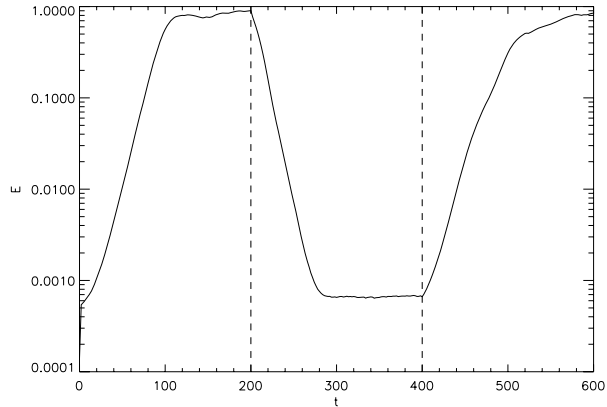}\\
\end{center}
\caption{$E(t)$ when the coupling is switched on (to a value of $\epsilon=0.1$) from $t=200$ to $t=400$ time units. The initial growth of $E$ is associated with the pattern formation process in both reactors, which starts from the uniform solution perturbed by noise. Remaining parameters as in Fig.\ \ref{fig:2}.}\label{fig:5}
\end{figure}

One might wonder whether the proposed coupling is general enough, or whether the reported phenomenon is robust. To elucidate (at least partially) on these questions, we have performed numerical experiments where the coupling is switched on only after two well-developed and independent NSS are formed in each reactor. The results are shown in Fig.\ \ref{fig:5}. The synchronization error decreases as soon as the coupling is switched on, a replication of system 1's NSS takes place after a transient, regardless of the initial condition on reactor 2. In particular, there is no need to stabilize the second reactor prior to synchronizing it with the first one. This fact shows explicitly the robustness of the observed phenomenon, and suggests that the attraction basin of the synchronization manifold ($a_2=a_1$, $b_2=b_1$) is large enough.

In the same way, if the coupling is only switched on in a part of the reactors, numerical simulations (not shown) indicate that (even for $\sigma_2=0$) a replication of the ``master structure'' takes place in the coupled region of the slave system, opening the possibility of local replication (see below).

\begin{figure}
\begin{center}
\includegraphics[width=3.2in,bb= 54pt 360pt 558pt 720pt]{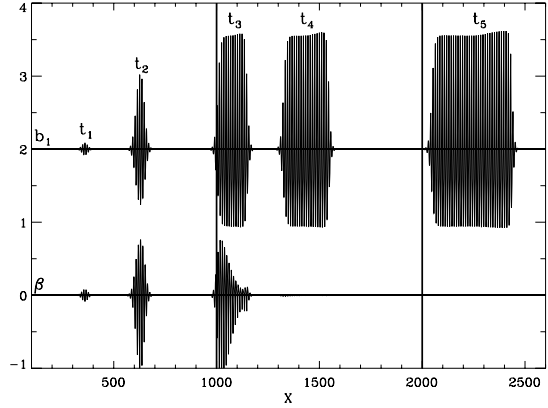}\\
\end{center}
\caption{Time evolution of a pulse-like initial condition for $\epsilon=0.1$, without noise ($\sigma_{1,2}=0$). The coupling is switched on between the vertical bold lines ($1000\leq x\leq2000$). Shown are the $b_1$ and $\beta$ fields for $t_1=40$, $t_2=70$, $t_3=120$, $t_4=160$ and $t_5=250$. Note that already for $t_4$, the deterministic structure in system $1$ is fully replicated in system $2$, and they remain synchronized even without coupling. The other parameters are $\mu=2.0$ and $\phi=9.5$.}\label{fig:6}
\end{figure}

\begin{figure}
\begin{center}
\includegraphics[width=3.2in,bb= 54pt 360pt 558pt 720pt]{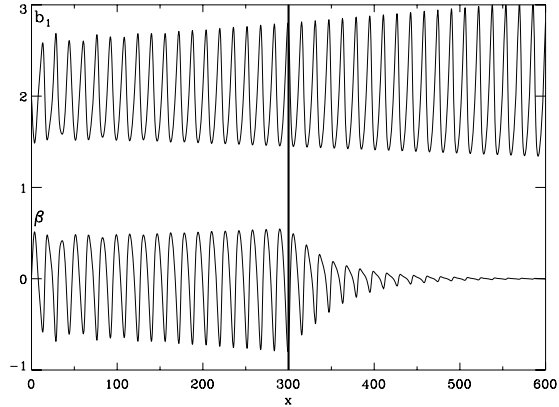}\\
\end{center}
\caption{Synchronization of an externally forced deterministic structure. We show a snapshot of the $b_1$ and $\beta$ fields. The left b.c.\ on system 1 $[b_1(x=0,t)=\mu+0.5\,\sin(4.71\,t)]$ generates a traveling structure. The coupling ($\epsilon=0.2$) is switched on at the right of the vertical bold line ($300\leq x$) and the synchronization induced by coupling results evident. Remaining parameters as in Fig.\ \ref{fig:6}.}\label{fig:7}
\end{figure}

We have also explored the synchronization of deterministic non-equilibrium structures (i.e, no noise in both systems). In particular, we have considered two cases of pattern generation: (a) structures generated by the time evolution of an inhomogeneous initial condition (Fig.\ \ref{fig:6}); and (b) structures generated by an oscillatory external forcing (Fig.\ \ref{fig:7}). In both cases, the numerical results (discused in the captions) clearly confirm the robustness of the proposed mechanism of synchronization.
\section{Conclusions}\label{sec:concl}
By coupling unidirectionally (but otherwise linearly) corresponding points of two samples of the convectively unstable system under study, complete synchronization of macroscopic structures has been achieved, both for deterministic and stochastic dynamics. Figure \ref{fig:5} suggests that the synchronization attractor is very extended and a full replication of the structures is to be expected under very general conditions. This is a strong indication that the synchronization manifold ($a_2=a_1$, $b_2=b_1$) is at least linearly stable. A complete stability analysis of the synchronization manifold will be published elsewhere \cite{IDB}. We remark that the coupling synchronizes completely both systems (after a transient) regardless of the initial condition in the ``slave'' system. Even more, the coupling may be defined in part of the system's extension. Thus, for $\sigma_2=0$, a (synchronized) structure arises in the slave system \emph{because} of the coupling.

We expect the phenomenon to have technological applications in the control of differential-flow chemical reactors, and eventually in the case where the convective structures in the master system are not noise-sustained ones, but carry useful information.

\end{document}